\documentclass[reprint,amsmath,amssymb,aps,prb,floatfix]{revtex4-2}

\usepackage{graphicx}
\usepackage{dcolumn}
\usepackage{bm}

\begin{document}


\title{Spin and charge fluctuations in the two-band Hubbard model}

\author{Alexei Sherman}
\affiliation{Institute of Physics, University of Tartu, 1 W. Ostwaldi Street, 50411 Tartu, Estonia}%

\date{\today}

\begin{abstract}
A model of CuO$_2$ planes of cuprate perovskites, containing $d_{x^2-y^2}$ copper orbitals and symmetric combinations of oxygen $p_\sigma$ orbitals, is investigated using the strong coupling diagram technique. This approach allows one to take into account the interactions of carriers with spin and charge fluctuations of all ranges. Derived equations for Green's function are self-consistently solved for the set of parameters corresponding to hole- and electron-doped cuprates. It is shown that the mentioned interactions lead to the appearance of spin polarons -- bound states of carriers with spin excitations, which show themselves as sharp peaks of the density of states and spectral functions at the Fermi level. Hole and electron doping are strongly asymmetric. This, in particular, manifests itself in the antiferromagnetic response for the electron-doped case and in an incommensurate magnetic ordering for hole doping. In the latter case, the incommensurability parameter grows with doping. The double occupancy shows that the electron-doped system retains strong correlations up to the concentration 0.23, while for hole doping the correlations decay rapidly. These results are in agreement with experimental observations in cuprates.
\end{abstract}

\maketitle


\section{Introduction}
The three-band Hubbard model contains a minimal set of states, which is necessary for the description of CuO$_2$ planes of cuprate high-$T_c$ superconductors -- a copper $3d_{x^2-y^2}$ and two oxygen $2p_\sigma$ orbitals per unit cell \cite{Zaanen,Emery,Varma}. Previously, the model was investigated by different methods including exact diagonalization of small clusters \cite{Horsch89,Horsch90}, Monte Carlo simulations \cite{Dopf,Scalettar}, the dynamic cluster approximation \cite{Macridin}, dynamic mean-field approximation (DMFT) \cite{Weber08,Medici,Weber10,Wang}, variational cluster approach \cite{Arrigoni}, density-matrix-renormalization-group calculations \cite{White} and the strong coupling diagram technique (SCDT) using two lowest orders of the series expansion \cite{Sherman16}. These works have shown that the model exhibits a number of the basic magnetic and single-particle spectral properties that are seen in the cuprates. They demonstrated also that the low-frequency part of the model spectrum has some similarity with the spectrum of the one-band Hubbard model.

The mentioned works either did not consider interactions of carriers with spin and charge fluctuations or took into account only their short-range part. In the present work, we take into consideration the fluctuations of all ranges using the SCDT \cite{Vladimir,Metzner,Pairault,Sherman18,Sherman19}. In this approach, Green's functions are calculated using series expansions in powers of the carrier intersite hopping term of the Hamiltonian. Terms of the series are products of hopping constants and on-site cumulants of carrier operators. The linked-cluster theorem is valid, and partial summations are allowed in this diagram technique (the concise description of the approach can be found in Ref.~\cite{Sherman16}). In the ladder approximation, the interactions of carriers with charge and spin fluctuations are described by diagrams with ladder inserts. If these ladders are constructed from renormalized hopping lines and second-order cumulants, ladders of all lengths can be summed \cite{Sherman18}. Thereby, fluctuations of all ranges are taken into account in an infinite crystal. In Refs.~\cite{Sherman18,Sherman19}, it was shown that the spectral functions, magnetic susceptibility, double occupancy and squared site spin calculated in the one-band Hubbard model using the SCDT are in good agreement with results of exact diagonalizations, Monte Carlo simulations, numerical linked-cluster expansions and experiments with ultracold fermionic atoms in two-dimensional (2D) optical lattices in wide ranges of repulsions, temperatures, and concentrations. This fact gives grounds to believe that this approach will be equally useful for investigating many-band Hubbard and Hubbard-Hund models.

To simplify somewhat the consideration of the present work, we set the oxygen-oxygen hopping constant and intersite repulsion to zero. In this case, oxygen states formed from the antisymmetric site combinations do not interact with copper and symmetric oxygen states and can be omitted. For this two-band model, the SCDT leads to a system of equations, which can be solved by iteration. Calculations are carried out for the entire range of hole concentrations $0\leq x\leq 4$ using the Hubbard-I approximation. In this approach, we found that there are five regions of the chemical potential, in which the spectrum and states contributing to it are fundamentally different. The influence of spin and charge fluctuations on the spectrum is investigated for hole concentrations $0.77\lesssim x\lesssim 1.28$ topical for cuprates. At low temperatures, the interaction of holes with spin and charge fluctuations leads to the appearance of spin polarons -- bound states of holes and spin excitations. The excitations manifest themselves as sharp peaks at the Fermi level (FL) in spectral bands. Similar spin polarons were earlier observed in the one-band Hubbard model \cite{Sherman19,Sherman19a}. In their appearance, the spin-polaron peaks resemble the DMFT quasiparticle peaks, which originate from the Abrikosov-Suhl resonances of the Anderson impurity model \cite{Georges,Hewson}. We emphasize that in the SCDT, all calculations are performed strictly within the framework of the 2D Hubbard model, and polarons are bound states of its excitations.

Using the obtained Green's functions and vertices, we calculate spin $\chi^{\rm sp}$ and charge $\chi^{\rm ch}$ susceptibilities, double occupancy, and the square of the site spin. These quantities demonstrate a strong asymmetry of electron $x<1$ and hole $x>1$ doping. For electron doping, the zero-frequency $\chi^{\rm sp}$ on copper sites is peaked at the antiferromagnetic wave vector ${\bf k}=(\pi,\pi)$ up to the electron concentration 0.23 (the distance between copper sites is set as the unit of length). For hole doping, the magnetic response is incommensurate, and the incommensurability parameter grows with doping. Qualitatively, these results reproduce the known peculiarities of the magnetic response in the hole- and electron-doped cuprate perovskites \cite{Armitage,Fujita,Wilson}. At $x=1$, the zero-frequency magnetic susceptibility at $(\pi,\pi)$ can serve as the measure of the proximity to the long-range antiferromagnetic ordering. In the present model, it is noticeably smaller than the value in the one-band Hubbard model for comparable parameters. Thus, spin correlations fall off by the $p$-$d$ hybridization. The double occupancy indicates the strength of charge correlations. For the used parameters, it is small at $x=1$ and remains practically the same at electron doping up to the concentration 0.23. Hence the considered system retains strong correlations, even being heavily doped by electrons. In contrast, at hole doping, the double occupancy grows rapidly, pointing to the significant decay of correlations. A similar conclusion was made earlier analyzing experimental results \cite{Armitage}. Both types of doping lead to the decrease of the squared site spin. However, on the electron side, the decrease is more pronounced due to the combined action of changes in the double occupancy and the hole number.

The paper is organized as follows. In Sec.~II, the two-band Hubbard model and the SCDT are introduced. The Hubbard-I approximation is discussed in Sec.~III. Its results give an overview of spectral shapes in different regions of the chemical potential and serve as the starting point of forthcoming iterations. Equations for Green's function, which take into account interactions of holes with spin and charge fluctuations, the discussion of the calculation procedure, obtained densities of states (DOS), and spectral functions are given Sec.~IV. Results on the magnetic susceptibility, double occupancy, and square of site spin are presented in Sec.~V. The conclusions of this study are reviewed in Sec.~VI.

\section{Model and SCDT method}
The Hamiltonian of the model reads \cite{Zaanen,Emery,Varma}
\begin{eqnarray}\label{3bands}
 H&=&(\Delta-\mu)\sum_{\bf lz\sigma}p^\dagger_{\bf l+z,\sigma}p_{\bf l+z,\sigma}+
 \frac{U}{2}\sum_{\bf l\sigma}n_{\bf l\sigma}n_{\bf l,-\sigma}\nonumber\\
 &+&t\sum_{{\bf lz}s\sigma}s\left(d^\dagger_{\bf l\sigma}p_{{\bf l}+s{\bf z},\sigma}+
 p^\dagger_{{\bf l}+s{\bf z},\sigma}d_{\bf l\sigma}\right)-\mu\sum_{\bf l\sigma}n_{\bf l\sigma},
\end{eqnarray}
where $\Delta$ is the energy difference between $p_x$, $p_y$ and $d_{x^2-y^2}$ levels of oxygen and copper ions, $\mu$ is the chemical potential, $d^\dagger_{\bf l\sigma}$ and $d_{\bf l\sigma}$ are creation and annihilation operators of {\em holes} on copper ions, {\bf l} labels sites of a square lattice, $\sigma=\pm 1$ is the spin projection, $n_{\bf l\sigma}=d^\dagger_{\bf l\sigma}d_{\bf l\sigma}$, $p^\dagger_{\bf l+z,\sigma}$ and $p_{\bf l+z,\sigma}$ are creation and annihilation operators of holes on oxygen ions located halfway between copper ions, ${\bf z}={\bf x}/2$, ${\bf y}/2$, and {\bf x} and {\bf y} are the elementary translations of the copper lattice, $U$ is the energy of the Hubbard repulsion between two holes occupying the same copper site, $t$ is the copper-oxygen hopping energy, and $s=\pm 1$ in the next to the last term takes into account phases of the respective wave functions. To simplify further consideration, hopping terms between different oxygen orbitals, Coulomb repulsion terms between holes on neighboring oxygen sites and oxygen and copper sites are neglected in the present work. Due to the low hole concentration on oxygen sites, the mentioned Coulomb terms can be thought of as included at the mean-field level, which modifies $\Delta$. In more elaborate calculations, the inter-oxygen hopping can be included into the consideration along the lines discussed below.

After the Fourier transformation,
\begin{equation*}
d_{\bf k\sigma}=\frac{1}{\sqrt{N}}\sum_{\bf l}{\rm e}^{{\rm i}\bf kl}d_{\bf l\sigma}, \;
p_{\bf kz\sigma}=\frac{1}{\sqrt{N}}\sum_{\bf l}{\rm e}^{{\rm i}\bf k(l+z)}p_{\bf l+z,\sigma},
\end{equation*}
where $N$ is the number of copper sites and {\bf k} is the 2D wave vector it is convenient to introduce the new oxygen operators corresponding to symmetric and antisymmetric combinations of initial operators
\begin{eqnarray*}
\phi_{\bf k\sigma}&=&-{\rm i}\alpha_{\bf k}^{-1}\left[\sin({\bf kx}/2)p_{\bf kx\sigma}+\sin({\bf ky}/2)p_{\bf ky\sigma}\right], \\
\psi_{\bf k\sigma}&=&-{\rm i}\alpha_{\bf k}^{-1}\left[\sin({\bf ky}/2)p_{\bf kx\sigma}-\sin({\bf kx}/2)p_{\bf ky\sigma}\right]
\end{eqnarray*}
with $\alpha_{\bf k}=\left[\sin({\bf kx}/2)^2+\sin({\bf ky}/2)^2\right]^{1/2}$. The new ope\-ra\-tors satisfy the usual anticommutation relations. Substituting these operators into the Hamiltonian (\ref{3bands}), one can see that the system splits into two subsystems, one of which contains uncorrelated excitations described by $\psi_{\bf k\sigma}$ and another is formed by correlated excitations connected with $d_{\bf k\sigma}$ and $\phi_{\bf k\sigma}$. The contribution of states of the first subsystem is easily calculated, and it will not be considered here. After discarding the respective term and the reverse transformation to the site representation, the Hamiltonian reads
\begin{eqnarray}\label{2bands}
H=\sum_{\bf l} H_{\bf l}+\sum_{\bf ll'}\sum_{ii'\sigma}t_{{\bf l}i,{\bf l'}i'}a^\dagger_{{\bf l'}i'\sigma}a_{{\bf l}i\sigma},
\end{eqnarray}
where
\begin{eqnarray}\label{local}
H_{\bf l}&=&\sum_\sigma\bigg[(\Delta-\mu)n_{{\bf l}2\sigma}-\mu n_{{\bf l}1\sigma} +\frac{U}{2}n_{{\bf l}1\sigma}n_{{\bf l}1,-\sigma}\nonumber\\
&+&2t\alpha_{\bf 0}\left(a^\dagger_{{\bf l}1\sigma}a_{{\bf l}2\sigma}+a^\dagger_{{\bf l}2\sigma}a_{{\bf l}1\sigma}\right)\bigg],
\end{eqnarray}
$a_{{\bf l}1\sigma}=d_{{\bf l}\sigma}$, $a_{{\bf l}2\sigma}=\phi_{{\bf l}\sigma}$, $n_{{\bf l}i\sigma}=a^\dagger_{{\bf l}i\sigma}a_{{\bf l}i\sigma}$, $i=1,\,2$,
\begin{equation}\label{hopping}
t_{{\bf l}i,{\bf l'}i'}=2t(\alpha_{\bf l-l'}-\alpha_{\bf 0}\delta_{\bf ll'})(1-\delta_{ii'}),
\end{equation}
$\alpha_{\bf l}=N^{-1}\sum_{\bf k}\exp({\rm i}{\bf kl})\alpha_{\bf k}$, $\alpha_{\bf 0}=N^{-1}\sum_{\bf k}\alpha_{\bf k}\approx 0.958$.

Terms collected in the Hamiltonian (\ref{local}) contain operators belonging to one site, while the second term in Eq.~(\ref{2bands}) describes intersite hopping. The Hamiltonian $H_{\bf l}$ can be easily diagonalized, and its 16 eigenvalues and eigenstates will be denoted as $E_\lambda$ and $|\lambda\rangle$.

We shall calculate the following one-particle Green's functions: $G_{i'i}({\bf l'}\tau';{\bf l}\tau)=\langle{\cal T}\bar{a}_{{\bf l'}i'\sigma}(\tau')a_{{\bf l}i\sigma}(\tau)\rangle$, where ${\cal T}$ is the chronological operator, the thermodynamic averaging and time dependencies are determined by the Hamiltonian (\ref{2bands}),
\begin{equation*}
\bar{a}_{{\bf l}i}(\tau)={\rm e}^{H\tau}a^\dagger_{{\bf l}i}{\rm e}^{-H\tau}.
\end{equation*}
For this purpose we use the SCDT, in which the series expansion is performed over powers of the second term of the Hamiltonian (\ref{2bands}). Terms of this expansion are products of hopping integrals $t_{{\bf l}i,{\bf l'}i'}$ (\ref{hopping}) and on-site cumulants \cite{Kubo} of operators $a^\dagger_{i\sigma}$ and $a_{i\sigma}$. In the present case, besides spin and time variables, the cumulants depend on indices $i=1,\,2$ distinguishing two site states -- $d$ and $\phi$. Each term of the expansion can be represented graphically as a diagram, in which hopping integrals are depicted by directed lines and cumulants by circles with the number of outgoing (ingoing) lines corresponding to the cumulant order.

\begin{figure}[t]
\centerline{\resizebox{0.99\columnwidth}{!}{\includegraphics{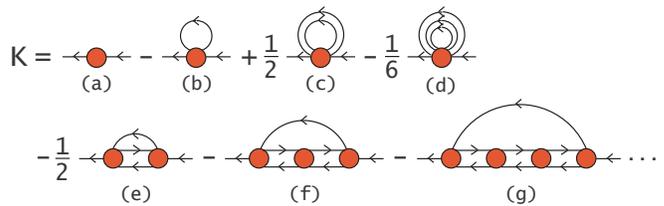}}}
\caption{Diagrams of several lowest orders in the SCDT expansion for ${\bf K}({\bf k},j)$.} \label{Fig1}
\end{figure}

As in the usual diagram technique with the series expansion over powers of an interaction \cite{Abrikosov}, the diagrams can be divided into reducible and irreducible. The latter cannot be separated into two disconnected parts by cutting a hopping line. If the sum of all irreducible diagrams -- the irreducible part -- is denoted by {\bf K} the Fourier transform of the Green's function can be represented as
\begin{equation}\label{Larkin}
{\bf G}({\bf k},j)={\bf K}({\bf k},j)[{\bf 1}-{\bf t_k}{\bf K}({\bf k},j)]^{-1},
\end{equation}
where $j$ is an integer defining the Matsubara frequency $\omega_j=(2j-1)\pi T$ with the temperature $T$. In Eq.~(\ref{Larkin}), we use matrix notations for quantities depending on two indices $i$; {\bf 1} is the $2\times 2$ unit matrix, and
$${\bf t_k}=2t(\alpha_{\bf k}-\alpha_{\bf 0})\left(\begin{array}{cc}
                                                     0 & 1 \\
                                                     1 & 0
                                                   \end{array}\right).$$
Equation (\ref{Larkin}) is the direct matrix generalization of the respective SCDT equation for the one-band Hubbard model \cite{Sherman18}. Diagrams look similarly also, as can be seen from the comparison of Fig.~\ref{Fig1} with diagrams in Refs.~\cite{Sherman18,Sherman19}. However, in spite of the formal resemblance, there is an essential difference -- in the case of the one-band Hubbard model, hopping lines and cumulants are scalars, while in Fig.~\ref{Fig1} they are matrices and tensors of indices $i$. Summations are carried out over these indices of internal lines, as well as over frequencies and spin indices.

In its form, Eq.~(\ref{Larkin}) is similar to the equation of the cluster perturbation theory \cite{Senechal} used in the consideration of the one-band Hubbard model. In this theory, the role of our local system (\ref{local}) is played by a small cluster, $t_{{\bf l}i,{\bf l'}i'}$ describes an intercluster hopping and the irreducible part $K$ is approximated by the first-order cumulant.

\section{The Hubbard-I approximation}
In this section, we consider an approach, which is equivalent to the Hubbard-I approximation \cite{Hubbard} of the one-band model. In the SCDT, it is obtained by approximating the full irreducible part with the first term of its power expansion -- the first-order cumulant \cite{Vladimir}. Thus, we set ${\bf K}={\bf C}^{(1)}$, where, in our case, the first-order cumulant $C^{(1)}_{i'i}(\tau',\tau)=\langle{\cal T}\bar{a}_{i'\sigma}(\tau') a_{i\sigma}(\tau)\rangle_0$. Following the SCDT idea, the thermodynamic averaging and time dependence are determined by the site Hamiltonian (\ref{local}), which is indicated by the subscript 0. This approximation allows us to obtain an overview of the DOS in the entire range of hole concentrations. Besides, the Hubbard-I results will be used as starting values in the iteration procedure of the next section. The calculation of ${\bf C}^{(1)}$ can be easily performed in the representation of eigenvectors of $H_{\bf l}$,
\begin{equation}\label{lambdarepr}
a_{i\sigma}=\sum_{\lambda\lambda'}\langle\lambda|a_{i\sigma}|\lambda'
\rangle X^{\lambda\lambda'},
\end{equation}
where the Hubbard operator $X^{\lambda\lambda'}=|\lambda\rangle\langle \lambda'|$. The result reads
\begin{equation}\label{C1}
C^{(1)}_{i'i}(j)=\frac{1}{Z}\sum_{\lambda\lambda'}
\frac{{\rm e}^{-\beta E_\lambda}+{\rm e}^{-\beta E_{\lambda'}}}{{\rm i}\omega_j+E_\lambda-E_{\lambda'}} \langle\lambda|a_{i\sigma}|\lambda'\rangle\langle\lambda'|a^\dagger_{i'\sigma}
|\lambda\rangle
\end{equation}
with the partition function $Z=\sum_\lambda\exp(-\beta E_\lambda)$ and $\beta=1/T$.

\begin{figure*}[htb]
\centerline{\resizebox{1.5\columnwidth}{!}{\includegraphics{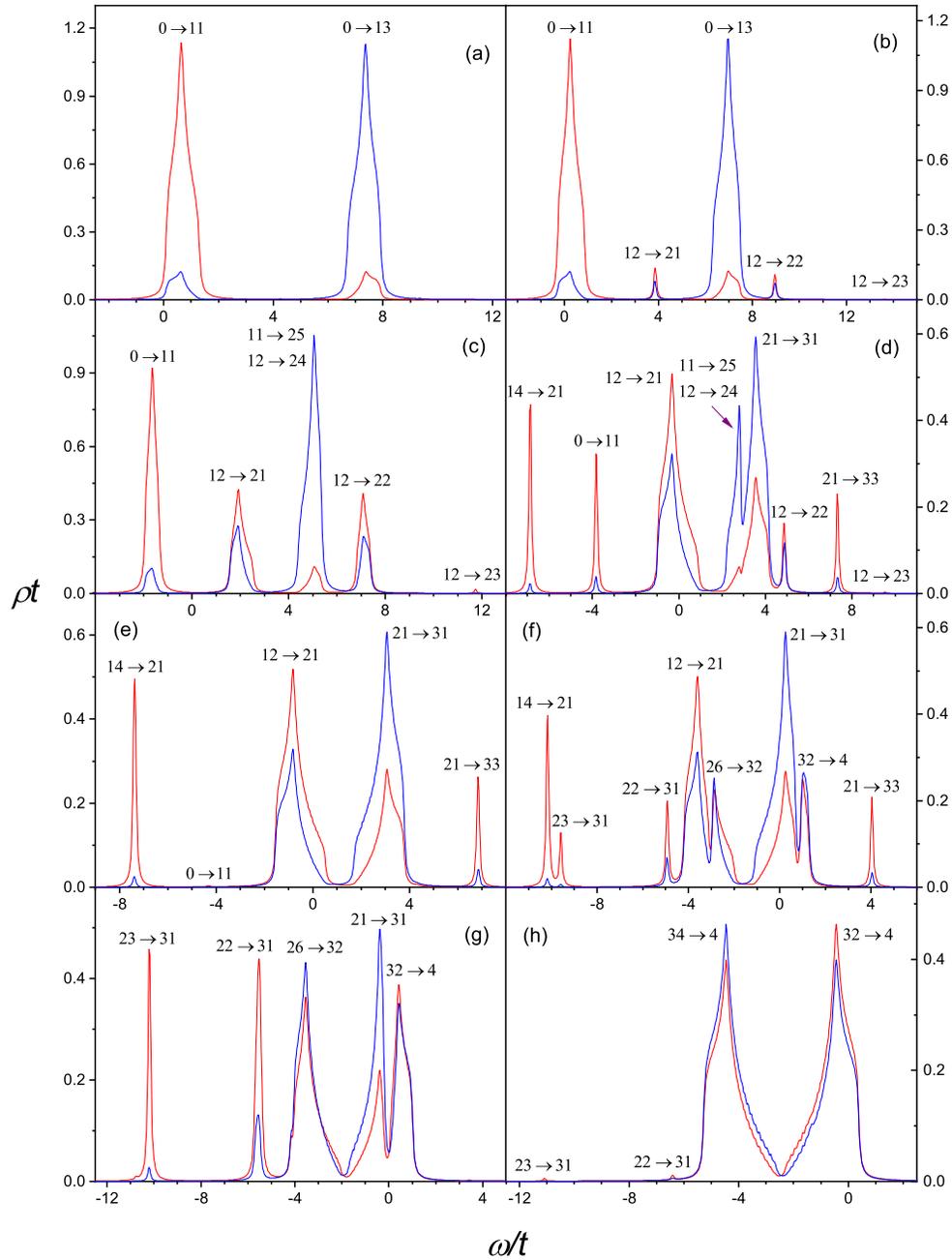}}}
\caption{The evolution of the densities of states on copper (red curves) and oxygen (blue curves) sites with the change of the chemical potential. $U=5.7t$, $\Delta=5.4t$, $T=0.1t$ and $\mu=-1.3t$ (the hole concentration $x\approx 0.15$, a), $-0.9t$ ($x\approx 0.50$, b), $t$ ($x\approx 1.01$, c), $3.2t$ ($x\approx 1.45$, d), $3.7t$ ($x\approx 1.79$, e), $6.5t$ ($x\approx 2.59$, f), $7.1t$ ($x\approx 2.97$, g), $8t$ ($x\approx 3.56$, h). The transitions in the local Hamiltonian (\protect\ref{local}), with which spectral bands are connected, are also shown.} \label{Fig2}
\end{figure*}
Bands in spectra obtained from Eqs.~(\ref{Larkin}) and (\ref{C1}) are related to transitions between eigenstates of the local Hamiltonian (\ref{local}). For their description, we shall characterize the eigenstates by the number of holes in them, $n$, and, in the case of several eigenstates with the same $n$, by the second index $k$. In the cases $n=1$ and 3 there are two degenerate lower states, which are denoted by $k=1$ and 2, and two higher states with $k=3$ and 4. Thus, the last-mentioned eigenvector is specified as $|34\rangle$. There are six states with $n=2$, three singlets designated by $k=1-3$ and three degenerate triplets with $k=4-6$. The lowest of singlets, $|21\rangle$, is the Zhang-Rice state \cite{Zhang}.

The eigenenergy $E_\lambda$ depends on $\mu$ through the term $-\mu n$. Therefore, as the chemical potential varies, states with different $n$ become alternately the ground state of the Hamiltonian (\ref{local}). Due to the Boltzmann factors $\exp(-\beta E_\lambda)$ in Eq.~(\ref{C1}), at low temperatures, this state (or states, in the case of degeneracy) and states obtained from it by the creation or annihilation of one hole make the main contribution to the cumulant and, through Eq.~(\ref{Larkin}), to spectral functions and DOS. Thus, spectra are qualitatively changed when $\mu$ transfers between regions with different ground states. There are five such regions in accord with five possible values of $n$.
We denote these regions by Roman numerals. Hereinafter we use the following parameters, which were suggested in earlier works \cite{Weber08,Weber10,Wang} for cuprates: $U=5.7t$ and $\Delta=5.4t$. For these parameters, the mentioned regions correspond to the following ranges of the chemical potential: $\mu\lesssim -0.61t$ (I), $-0.61t\lesssim\mu\lesssim 2.9t$ (II), $2.9t\lesssim\mu\lesssim 6.7t$ (III), $6.7t\lesssim\mu\lesssim 7.5t$ (IV), and $7.5t\lesssim\mu$ (V). In these ranges, states $|0\rangle$, $|11\rangle$ and $|12\rangle$, $|21\rangle$, $|31\rangle$ and $|32\rangle$, and $|4\rangle$ are the lowest, respectively. The boundary domain between the two regions has a width of several $T$. We shall denote it by numerals of regions it separates, for example, I-II.

The DOSs on copper and oxygen sites,
\begin{equation*}
\rho_i(\omega)=-(N\pi)^{-1}\sum_{\bf k}{\rm Im}\,G_{ii}({\bf k},\omega),
\end{equation*}
calculated for ${\bf K}({\bf k},\omega)={\bf C}^{(1)}(\omega)$ are shown in Fig.~\ref{Fig2}. Panels (a) to (h) demonstrate DOSs in the above-mentioned regions I, I-II, II, II-III, III, III-IV, IV, and V, respectively. The transitions between states of the local Hamiltonian, which are responsible for bands in Fig.~\ref{Fig2}, are also shown there. As stated above, these states and DOS shapes vary considerably in different regions. In boundary domains, spectra demonstrate features inherent in both neighbor regions. The sum $x=x_1+x_2$ of the hole concentrations on copper ($x_1$) and oxygen ($x_2$) sites,
\begin{equation*}
x_i=\int_{-\infty}^{\infty}\frac{\rho_i(\omega){\rm d}\omega}{\exp(\beta\omega)+1},
\end{equation*}
is indicated in the figure captions. The case shown in panel (c) corresponds to the situation, which is similar to the half-filled one-band Hubbard model. The band caused by the transition $|0\rangle\rightarrow|11\rangle$ corresponds to the lower Hubbard band, and the band connected with the transition to the Zhang-Rice singlet, $|12\rangle\rightarrow|21\rangle$, complies with the upper Hubbard band. Bands with higher energies owe their emergence to transitions to triplets and higher-energy singlets.

\begin{figure}[t]
\centerline{\resizebox{0.95\columnwidth}{!}{\includegraphics{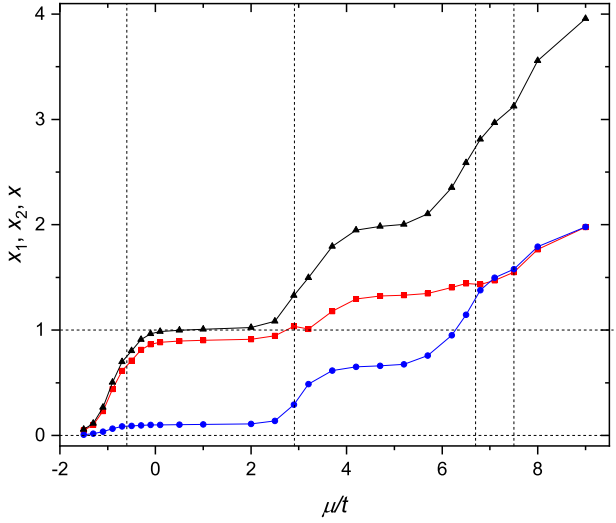}}}
\caption{Hole concentrations on copper (red curve and squares) and oxygen (blue curve and circles) sites and their sum (black curve and triangles) as functions of the chemical potential. Parameters are the same as in Fig.~\protect\ref{Fig2}. Vertical dashed lines separate the five regions of $\mu$ mentioned in the text.} \label{Fig3}
\end{figure}
Figure~\ref{Fig3} demonstrates the dependence of the hole concentrations on copper and oxygen sites and their sum on the chemical potential. The plateau at $x\approx 1$, which spans the range $0\lesssim\mu \lesssim 2t$ is connected with the charge-transfer gap between bands $|0\rangle\rightarrow|11\rangle$ and $|12\rangle\rightarrow|21\rangle$ in Fig.~\ref{Fig2}(c). The vicinity of this plateau will be considered in more detail in the next section taking into account more complicated processes.

\section{Spin and charge fluctuations}
In this section, we consider the influence of spin and charge fluctuations on hole spectra. In the ladder approximation, these processes are described by diagrams with ladder inserts, some of which are shown in the second row in Fig.~\ref{Fig1}. In these diagrams, as the irreducible four-leg vertex, we use the second-order cumulants of hole operators,
\begin{eqnarray*}
&&C^{(2)}_{i_1i_2i_3i_4}(\tau_1,\sigma_1;\tau_2,\sigma_2;\tau_3,\sigma_3;\tau_4,\sigma_4)\\
&&\quad=\big\langle{\cal T}\bar{a}_{\sigma_1i_1}(\tau_1)a_{\sigma_2i_2}(\tau_2) \bar{a}_{\sigma_3i_3}(\tau_3)a_{\sigma_4i_4}(\tau_4)\big\rangle_0\\
&&\quad\quad -C^{(1)}_{i_1i_2}(\tau_1,\tau_2)C^{(1)}_{i_3i_4}(\tau_3,\tau_4)\delta_{\sigma_1 \sigma_2}\delta_{\sigma_3\sigma_4}\\
&&\quad\quad+C^{(1)}_{i_1i_4}(\tau_1,\tau_4)C^{(1)}_{i_3i_2}(\tau_3, \tau_2)\delta_{\sigma_1 \sigma_4}\delta_{\sigma_3 \sigma_2}.
\end{eqnarray*}
We take into account ladder diagrams of all lengths, which allows us to consider fluctuations of all extensions. Besides, using the possibility of partial summation in the SCDT, we insert all possible two-leg diagrams in the internal lines of diagrams in Fig.~\ref{Fig1}. As a result, the bare hopping line ${\bf t_k}$ is substituted by the renormalized one described by the equation
\begin{equation}\label{theta}
\mbox{\boldmath $\theta$}({\bf k},j)={\bf t_k}+{\bf t_k}{\bf G}({\bf k},j){\bf t_k}.
\end{equation}

The irreducible part, which takes into account all these processes, reads
\begin{widetext}
\begin{eqnarray}\label{K}
&&K_{i'i}({\bf k},j)=C^{(1)}_{i'i}(j)+\frac{T^2}{2N}\sum_{{\bf k'}j'\nu}\sum_{i'_1i_1i'_2} \sum_{i_2i'_3i_3}\theta_{i_3i'_3}({\bf k'},j')
{\cal T}_{i_1i'_1i_2i'_2}({\bf k-k'},j+\nu,j'+\nu)\nonumber\\
&&\quad\times\bigg[\frac{3}{2}C^{(2)a}_{i'i_1i'_2i_3}(j,j+\nu,j'+\nu,j')
C^{(2)a}_{i'_1ii'_3i_2}(j+\nu,j,j',j'+\nu)
+\frac{1}{2}C^{(2)s}_{i'i_1i'_2i_3}(j,j+\nu,j'+\nu,j')
C^{(2)s}_{i'_1ii'_3i_2}(j+\nu,j,j',j'+\nu)\bigg] \nonumber\\
&&\quad-\frac{T}{N}\sum_{{\bf k'}j'}\sum_{i_1i'_1}\theta_{i_1i'_1}({\bf k'},j')
\bigg[\frac{3}{2}V^a_{i'ii'_1i_1}({\bf k-k'},j,j,j',j')
+\frac{1}{2}V^s_{i'ii'_1i_1}({\bf k-k'},j,j,j',j')\bigg].
\end{eqnarray}
\end{widetext}
In this equation, $C^{(2)a}$ and $C^{(2)s}$ are antisymmetrized and sym\-met\-ri\-zed over spin indices combinations of se\-cond-order cumulants,
\begin{eqnarray*}
&&C^{(2)a}_{i_1i'_1i_2i'_2}(j+\nu,j,j',j'+\nu)\\
&&\quad=\sum_{\sigma'}\sigma\sigma' C^{(2)}_{i_1i'_1i_2i'_2}(j+\nu,\sigma';j\sigma,j'\sigma,j'+\nu,\sigma'), \\
&&C^{(2)s}_{i_1i'_1i_2i'_2}(j+\nu,j,j',j'+\nu)\\
&&\quad=\sum_{\sigma'}C^{(2)}_{i_1i'_1i_2i'_2}(j+\nu,\sigma';j\sigma,j'\sigma,j'+\nu,\sigma').
\end{eqnarray*}
Due to the symmetry of the problem, they do not depend on the sign of the spin projection.
\begin{equation*}
{\cal T}_{i_1i'_1i_2i'_2}({\bf k},j,j')=\frac{1}{N}\sum_{\bf k'}\theta_{i_1i'_1}({\bf k+k'},j) \theta_{i_2i'_2}({\bf k'},j').
\end{equation*}
Quantities $V^a$ and $V^s$ are reducible four-leg vertices, which are analogously antisymmetrized and symmetrized over spin indices. $V^a$  satisfies the fol\-lo\-wing Bethe-Salpeter equation:
\begin{eqnarray}\label{BSE}
&&V^{a}_{i'ii'_1i_1}({\bf k},j+\nu,j,j',j'+\nu)\nonumber\\
&&\quad=C^{(2)a}_{i'ii'_1i_1}(j+\nu,j,j',j'+\nu)\nonumber\\
&&\quad +T\sum_{\nu'i_2i'_2}\sum_{i_3i'_3}C^{(2)a}_{i'i_3i'_2i_1}(j+\nu,j+\nu',j'+\nu',j'+\nu) \nonumber\\
&&\quad\times{\cal T}_{i_3i'_3i_2i'_2}({\bf k},j+\nu',j'+\nu')\nonumber\\
&&\quad\times V^{a}_{i'_3ii'_1i_2}({\bf k},j+\nu',j,j',j'+\nu').
\end{eqnarray}
The equation for $V^s$ looks similarly except that $C^{(2)a}$ is substituted with $C^{(2)s}$.

To make the above set of equations closed, we need the expression for the second-order cumulant. It can be derived using the representation (\ref{lambdarepr}) and the generalization of Wick's theorem for Hubbard operators \cite{Izyumov,Ovchinnikov}. The result reads
\begin{widetext}
\begin{eqnarray}\label{C2}
&&C^{(2)}_{i_1i_2i_3i_4}(j_1,\sigma_1;j_2,\sigma_1;j_3,\sigma_2;j_4,\sigma_2)=\frac{1}{Z} \sum_{\lambda_1\lambda_2}\sum_{\lambda_3\lambda_4}\langle\lambda_2|a^\dagger_{i_1\sigma_1}
|\lambda_1\rangle\langle\lambda_3|a_{i_2\sigma_1}|\lambda_2\rangle\langle\lambda_4|
a^\dagger_{i_3\sigma_2}|\lambda_3\rangle\langle\lambda_1|a_{i_4\sigma_2}|\lambda4\rangle
\nonumber\\
&&\quad\times\Big[g_{\lambda_1\lambda_4}(j_4)g_{\lambda_3\lambda_4}(j_3)
g_{\lambda_3\lambda_2}(j_2)\Big({\rm e}^{-\beta E_{\lambda_2}}+{\rm e}^{-\beta E_{\lambda_3}}\Big)-g_{\lambda_1\lambda_4}(j_4)g_{\lambda_3\lambda_4}(j_3)
g_{\lambda_4\lambda_2}(j_2-j_3)\Big({\rm e}^{-\beta E_{\lambda_2}}-{\rm e}^{-\beta E_{\lambda_4}}\Big)\nonumber\\
&&\quad+ g_{\lambda_1\lambda_4}(j_4)g_{\lambda_3\lambda_2}(j_2)
g_{\lambda_1\lambda_3}(j_4-j_3)\Big({\rm e}^{-\beta E_{\lambda_3}}-{\rm e}^{-\beta E_{\lambda_1}}\Big)+g_{\lambda_1\lambda_4}(j_4)g_{\lambda_3\lambda_2}(j_2)
g_{\lambda_1\lambda_2}(j_1)\Big({\rm e}^{-\beta E_{\lambda_2}}+{\rm e}^{-\beta E_{\lambda_1}}\Big)\Big]\nonumber\\
&&\quad+\langle\lambda_3|a^\dagger_{i_1\sigma_1}
|\lambda_1\rangle\langle\lambda_4|a_{i_2\sigma_1}|\lambda_2\rangle\langle\lambda_2|
a^\dagger_{i_3\sigma_2}|\lambda_3\rangle\langle\lambda_1|a_{i_4\sigma_2}|\lambda4\rangle
\nonumber\\
&&\quad\times\Big[-g_{\lambda_1\lambda_4}(j_4)g_{\lambda_3\lambda_2}(j_3)
g_{\lambda_4\lambda_2}(j_2)\Big({\rm e}^{-\beta E_{\lambda_2}}+{\rm e}^{-\beta E_{\lambda_4}}\Big)-g_{\lambda_1\lambda_4}(j_4)g_{\lambda_3\lambda_2}(j_3)
g_{\lambda_4\lambda_3}(j_2-j_3)\Big({\rm e}^{-\beta E_{\lambda_3}}-{\rm e}^{-\beta E_{\lambda_4}}\Big)\nonumber\\
&&\quad+ g_{\lambda_1\lambda_4}(j_4)g_{\lambda_3\lambda_2}(j_3)
g_{\lambda_1\lambda_2}(j_2+j_4-1)\Big({\rm e}^{-\beta E_{\lambda_2}}-{\rm e}^{-\beta E_{\lambda_1}}\Big)+g_{\lambda_1\lambda_4}(j_4)g_{\lambda_3\lambda_2}(j_3)
g_{\lambda_1\lambda_3}(j_1)\Big({\rm e}^{-\beta E_{\lambda_3}}+{\rm e}^{-\beta E_{\lambda_1}}\Big)\Big]\nonumber\\
&&\quad+\langle\lambda_4|a^\dagger_{i_1\sigma_1}
|\lambda_1\rangle\langle\lambda_3|a_{i_2\sigma_1}|\lambda_2\rangle\langle\lambda_1|
a^\dagger_{i_3\sigma_2}|\lambda_3\rangle\langle\lambda_2|a_{i_4\sigma_2}|\lambda_4\rangle
\nonumber\\
&&\quad\times\Big[g_{\lambda_2\lambda_4}(j_4)g_{\lambda_3\lambda_1}(j_3)
g_{\lambda_3\lambda_2}(j_2)\Big({\rm e}^{-\beta E_{\lambda_2}}+{\rm e}^{-\beta E_{\lambda_3}}\Big)-g_{\lambda_2\lambda_4}(j_4)g_{\lambda_3\lambda_1}(j_3)
g_{\lambda_1\lambda_2}(j_2-j_3)\Big({\rm e}^{-\beta E_{\lambda_2}}-{\rm e}^{-\beta E_{\lambda_1}}\Big)\nonumber\\
&&\quad+ g_{\lambda_2\lambda_4}(j_4)g_{\lambda_3\lambda_1}(j_3)
g_{\lambda_3\lambda_4}(j_2+j_4-1)\Big({\rm e}^{-\beta E_{\lambda_4}}-{\rm e}^{-\beta E_{\lambda_3}}\Big)-g_{\lambda_2\lambda_4}(j_4)g_{\lambda_3\lambda_1}(j_3)
g_{\lambda_1\lambda_4}(j_1)\Big({\rm e}^{-\beta E_{\lambda_4}}+{\rm e}^{-\beta E_{\lambda_1}}\Big)\Big]\nonumber\\
&&\quad+\langle\lambda_4|a^\dagger_{i_1\sigma_1}
|\lambda_1\rangle\langle\lambda_1|a_{i_2\sigma_1}|\lambda_2\rangle\langle\lambda_2|
a^\dagger_{i_3\sigma_2}|\lambda_3\rangle\langle\lambda_3|a_{i_4\sigma_2}|\lambda_4\rangle
\nonumber\\
&&\quad\times\Big[-g_{\lambda_3\lambda_4}(j_4)g_{\lambda_3\lambda_2}(j_3)
g_{\lambda_1\lambda_2}(j_2)\Big({\rm e}^{-\beta E_{\lambda_2}}+{\rm e}^{-\beta E_{\lambda_1}}\Big)-g_{\lambda_3\lambda_4}(j_4)g_{\lambda_3\lambda_2}(j_3)
g_{\lambda_1\lambda_3}(j_2-j_3)\Big({\rm e}^{-\beta E_{\lambda_3}}-{\rm e}^{-\beta E_{\lambda_1}}\Big)\nonumber\\
&&\quad+ g_{\lambda_3\lambda_4}(j_4)g_{\lambda_1\lambda_2}(j_2)
g_{\lambda_2\lambda_4}(j_4-j_3)\Big({\rm e}^{-\beta E_{\lambda_4}}-{\rm e}^{-\beta E_{\lambda_2}}\Big)-g_{\lambda_3\lambda_4}(j_4)g_{\lambda_1\lambda_2}(j_2)
g_{\lambda_1\lambda_4}(j_1)\Big({\rm e}^{-\beta E_{\lambda_4}}+{\rm e}^{-\beta E_{\lambda_1}}\Big)\Big]\nonumber\\
&&\quad+\langle\lambda_2|a^\dagger_{i_1\sigma_1}
|\lambda_1\rangle\langle\lambda_4|a_{i_2\sigma_1}|\lambda_2\rangle\langle\lambda_1|
a^\dagger_{i_3\sigma_2}|\lambda_3\rangle\langle\lambda_3|a_{i_4\sigma_2}|\lambda4\rangle
\nonumber\\
&&\quad\times\Big[-g_{\lambda_3\lambda_4}(j_4)g_{\lambda_3\lambda_1}(j_3)
g_{\lambda_1\lambda_2}(j_1)\Big({\rm e}^{-\beta E_{\lambda_2}}+{\rm e}^{-\beta E_{\lambda_1}}\Big)+g_{\lambda_3\lambda_4}(j_4)g_{\lambda_3\lambda_1}(j_3)
g_{\lambda_3\lambda_2}(j_2+j_4-1)\Big({\rm e}^{-\beta E_{\lambda_2}}-{\rm e}^{-\beta E_{\lambda_3}}\Big)\nonumber\\
&&\quad+ g_{\lambda_3\lambda_4}(j_4)g_{\lambda_4\lambda_2}(j_2)
g_{\lambda_1\lambda_4}(j_4-j_3)\Big({\rm e}^{-\beta E_{\lambda_4}}-{\rm e}^{-\beta E_{\lambda_1}}\Big)+g_{\lambda_3\lambda_4}(j_4)g_{\lambda_4\lambda_2}(j_2)
g_{\lambda_1\lambda_2}(j_1)\Big({\rm e}^{-\beta E_{\lambda_2}}+{\rm e}^{-\beta E_{\lambda_1}}\Big)\Big]\nonumber\\  
&&\quad+\langle\lambda_3|a^\dagger_{i_1\sigma_1}
|\lambda_1\rangle\langle\lambda_1|a_{i_2\sigma_1}|\lambda_2\rangle\langle\lambda_4|
a^\dagger_{i_3\sigma_2}|\lambda_3\rangle\langle\lambda_2|a_{i_4\sigma_2}|\lambda4\rangle
\nonumber\\
&&\quad\times\Big[g_{\lambda_2\lambda_4}(j_4)g_{\lambda_3\lambda_4}(j_3)
g_{\lambda_1\lambda_3}(j_1)\Big({\rm e}^{-\beta E_{\lambda_3}}+{\rm e}^{-\beta E_{\lambda_1}}\Big)+g_{\lambda_2\lambda_4}(j_4)g_{\lambda_3\lambda_4}(j_3)
g_{\lambda_1\lambda_4}(j_2+j_4-1)\Big({\rm e}^{-\beta E_{\lambda_4}}-{\rm e}^{-\beta E_{\lambda_1}}\Big)\nonumber\\
&&\quad+ g_{\lambda_2\lambda_4}(j_4)g_{\lambda_1\lambda_2}(j_2)
g_{\lambda_2\lambda_3}(j_4-j_3)\Big({\rm e}^{-\beta E_{\lambda_3}}-{\rm e}^{-\beta E_{\lambda_2}}\Big)-g_{\lambda_2\lambda_4}(j_4)g_{\lambda_1\lambda_2}(j_2)
g_{\lambda_1\lambda_3}(j_1)\Big({\rm e}^{-\beta E_{\lambda_3}}+{\rm e}^{-\beta E_{\lambda_1}}\Big)\Big]\nonumber\\
&&\quad-\beta\delta_{j_1j_2}C^{(1)}_{i_1i_2}(j_1)C^{(1)}_{i_3i_4}(j_3)+ \beta\delta_{j_1j_4}\delta_{\sigma_1\sigma_2}C^{(1)}_{i_1i_4}(j_1)C^{(1)}_{i_3i_2}(j_3),
\end{eqnarray}
\end{widetext}
where the energy conservation implies that $j_1+j_3=j_2+j_4$ and $g_{\lambda\lambda'}(j)=({\rm i}\omega_j+E_\lambda-E_{\lambda'})^{-1}$. In the latter expression, the Matsubara frequency may be either fermionic, $\omega_j=(2j-1)\pi T$, or bosonic, $\omega_j=2j\pi T$. The former case takes place when $g_{\lambda\lambda'}$ depends on only one $j$, the latter when it depends on the sum or difference of two such parameters. In this case, $g_{\lambda\lambda'}(j)$ diverges when $j=0$ and $E_\lambda=E_{\lambda'}$. However, such $g$ enter into Eq.~(\ref{C2}) together with differences of the respective Boltzmann factors. These products are finite in the mentioned conditions,
\begin{equation*}
g_{\lambda\lambda'}(j)\Big({\rm e}^{-\beta E_{\lambda'}}-{\rm e}^{-\beta E_\lambda}\Big) \stackrel{E_\lambda\rightarrow E_{\lambda'}}{\longrightarrow}{\rm e}^{-\beta E_\lambda} \beta\delta_{j0}.
\end{equation*}
It is worth noting that Eq.~(\ref{C2}) is the most general expression for the second-order cumulant, which is appropriate for any local Hamiltonian.

\begin{figure*}[t]
\centerline{\resizebox{1.5\columnwidth}{!}{\includegraphics{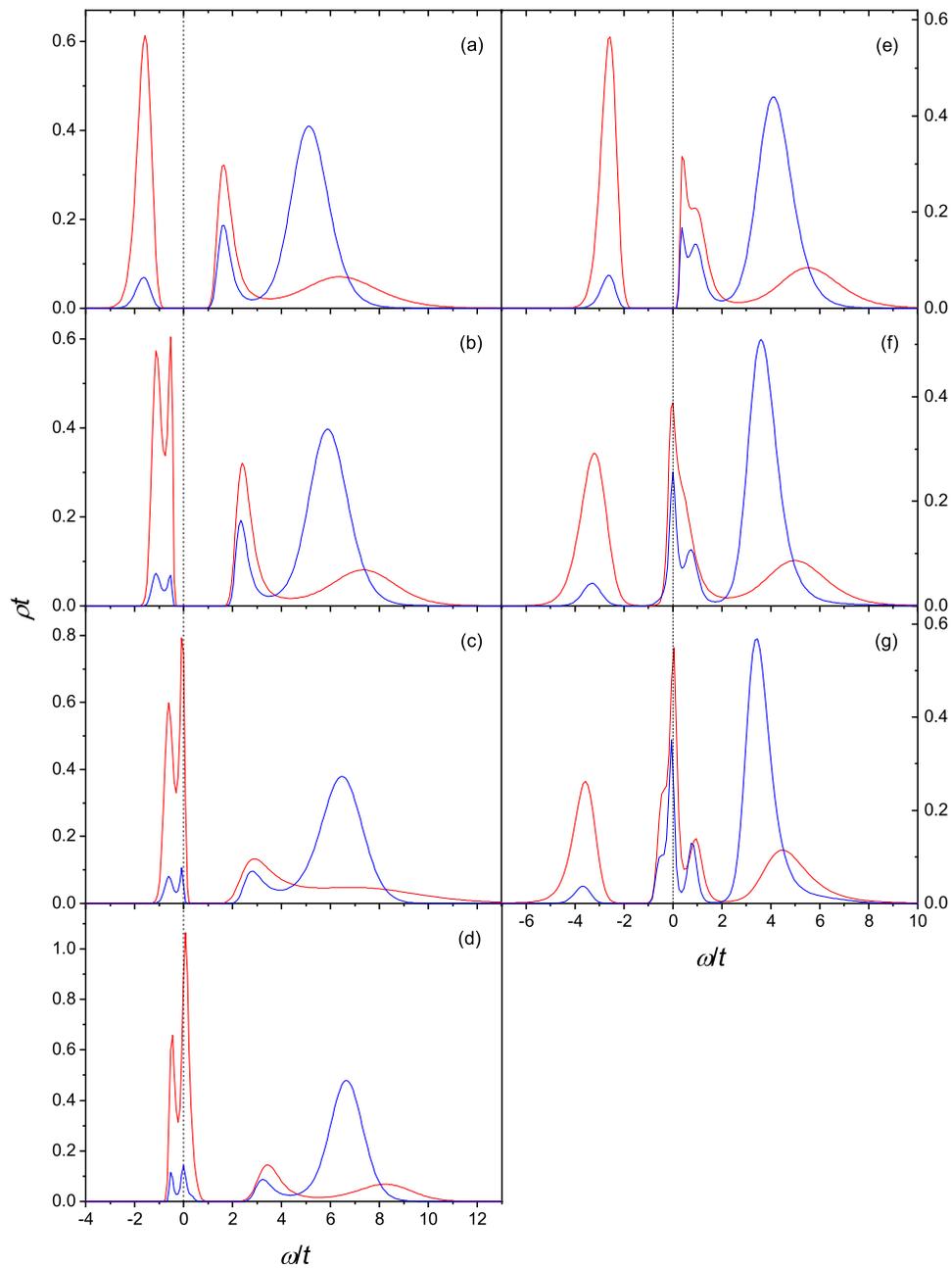}}}
\caption{The densities of states on copper (red curves) and oxygen (blue curves) sites, calculated taking into account interactions of holes with spin and charge fluctuations. $U=5.7t$, $\Delta=5.4t$, $T=0.12t$, and $\mu=t$ ($x=1$, a), $0.25t$ ($x=1$, b), $-0.25t$ ($x=0.93$, c), $-0.5t$ ($x=0.77$, d), $2t$ ($x=1.01$, e), $2.5t$ ($x=1.16$, f), $2.7t$ ($x=1.28$, g).} \label{Fig4}
\end{figure*}
Equations~(\ref{Larkin}), (\ref{C1})--(\ref{C2}) form a closed set, which can be solved by iteration for given values of $U$, $\Delta$, $T$, and $\mu$ expressed in units of $t$. The calculation consists of two stages. In the first stage, for a hole Green's function obtained in the previous step, the Bethe-Salpeter equations (\ref{BSE}) for the vertices $V^a$ and $V^s$ are solved. As starting values, the respective second-order cumulants $C^{(2)a}$ and $C^{(2)s}$ were used. This calculation stage is significantly simplified if notice that the matrix index of the linear system (\ref{BSE}) consists of only three variables -- $i'$, $i_1$, and $\nu$, while other variables -- ${\bf k}$, $j$, $j'$, $i$, and $i'_1$ -- are parameters. For the considered parameters, 7--10 iteration steps were enough to achieve convergence. In the second stage, the obtained vertices are used for calculating the hole Green's function from Eqs.~(\ref{K}) and (\ref{Larkin}). It is used for obtaining new vertices, and this cycle repeated until convergence. For the considered parameters, 10--20 cycles were necessary for this. In this iteration procedure, the result of the Hubbard-I approximation was used as the initial Green's function. No artificial broadening was introduced. The integration over wave vectors was approximated by the summation over the mesh of an 8$\times$8 lattice. It has nothing to do with the crystal finiteness, rather it is an approximate method of numerical integration, and the obtained results correspond to an infinite system.

The DOSs calculated using this procedure for hole concentrations near $x=1$ are shown in Fig.~\ref{Fig4}. To perform the analytic continuation from the obtained imaginary-frequency values to real-frequency ones, we used the maximum entropy method \cite{Press,Jarrell,Habershon}. Figures~\ref{Fig4}(a) and \ref{Fig2}(c) were calculated for the same parameters, with and without taking into account the interactions of holes with spin and charge fluctuations. For these parameters, the FL resides in the middle of the gap. The figures are rather similar. Interactions of holes with fluctuations lead to some broadening of the bands that results in merging the bands $|11\rangle\rightarrow|25\rangle$, $|12\rangle\rightarrow|24\rangle$ with $|12\rangle\rightarrow|22\rangle$.

\begin{figure}[t]
\centerline{\resizebox{0.99\columnwidth}{!}{\includegraphics{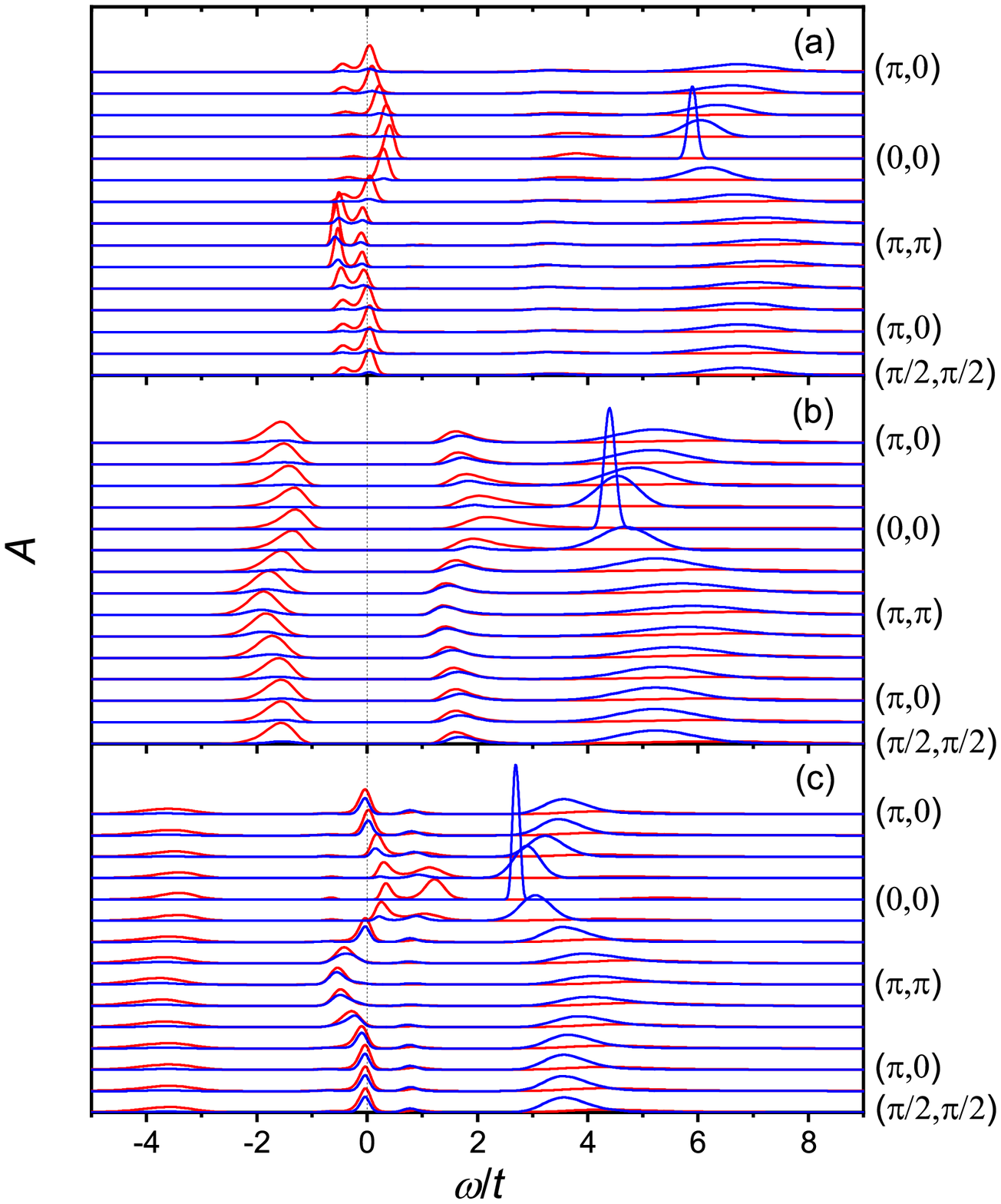}}}
\caption{Spectral functions $A({\bf k},\omega)$ on copper (red lines) and oxygen (blue lines) sites, at momenta on the symmetry lines of the Brillouin zone. The respective wave vectors are shown on the right ordinate. $U=5.7t$, $\Delta=5.4t$, $T=0.12t$, $\mu=-0.5t$ ($x=0.77$, a), $\mu=t$ ($x=1$, b), and $\mu=2.7t$ ($x=1.28$, c).}\label{Fig5}
\end{figure}
More significant changes in comparison with Hub\-bard-I results occur when the FL approaches the lower or upper band. Already before the crossing, a sharp peak appears in the respective band (panels b and e). After the FL enters the band, the peak is kept at the FL in a wide range of hole concentrations. Having regard to processes leading to the appearance of this peak and earlier results obtained in the one-band Hubbard \cite{Sherman19,Sherman19a} and the $t$-$J$ models (see, e.g., \cite{Schmitt,Martinez,Sherman94}), it can be connected with spin-polaron states. They are bound states of holes and spin excitations. In the mentioned simpler models, one can see that the band formed by these states has the width of the order of the exchange constant. The bandwidth is determined by the slower component of these composed excitations -- by the spin subsystem. Spin polarons and the associated peak in the DOS disappear with increasing temperature when spin excitations decay.

In its location and behavior, the spin-polaron peak resembles the DMFT resonance peak \cite{Georges}. It has to be underlined that in the SCDT, the peak was obtained strictly within the framework of the 2D Hubbard model, and excitations producing it belong to this model. In the DMFT, the resonant peak is a derivative of the Abrikosov-Suhl resonance of the Anderson impurity model. However, in spite of the significant difference between the approaches, the DOS in Fig.~\ref{Fig4}(g) and that calculated using the DMFT for close parameters in Ref.~\cite{Weber08} are similar. A slightly wider spectrum in the latter work can be related to a nonzero value of the oxygen-oxygen hopping.

Photoemission spectra of some crystals have peculiarities, which can be supposed to coincide with the spin-polaron peak. For example, maxima kept near the FL in wide ranges of doping were observed in electron- \cite{Armitage,Matsui} and hole-doped \cite{Damascelli} cuprates as well as in transition metal oxides \cite{Inoue}. In Ref.~\cite{Sherman19a}, it was shown that the dispersion of the band calculated in the framework of the one-band Hubbard model and associated with the peak coincides with the dispersion observed \cite{Matsui} in Nd$_{2-x}$Ce$_x$CuO$_4$ for comparable doping.

\begin{figure}[t]
\centerline{\resizebox{0.9\columnwidth}{!}{\includegraphics{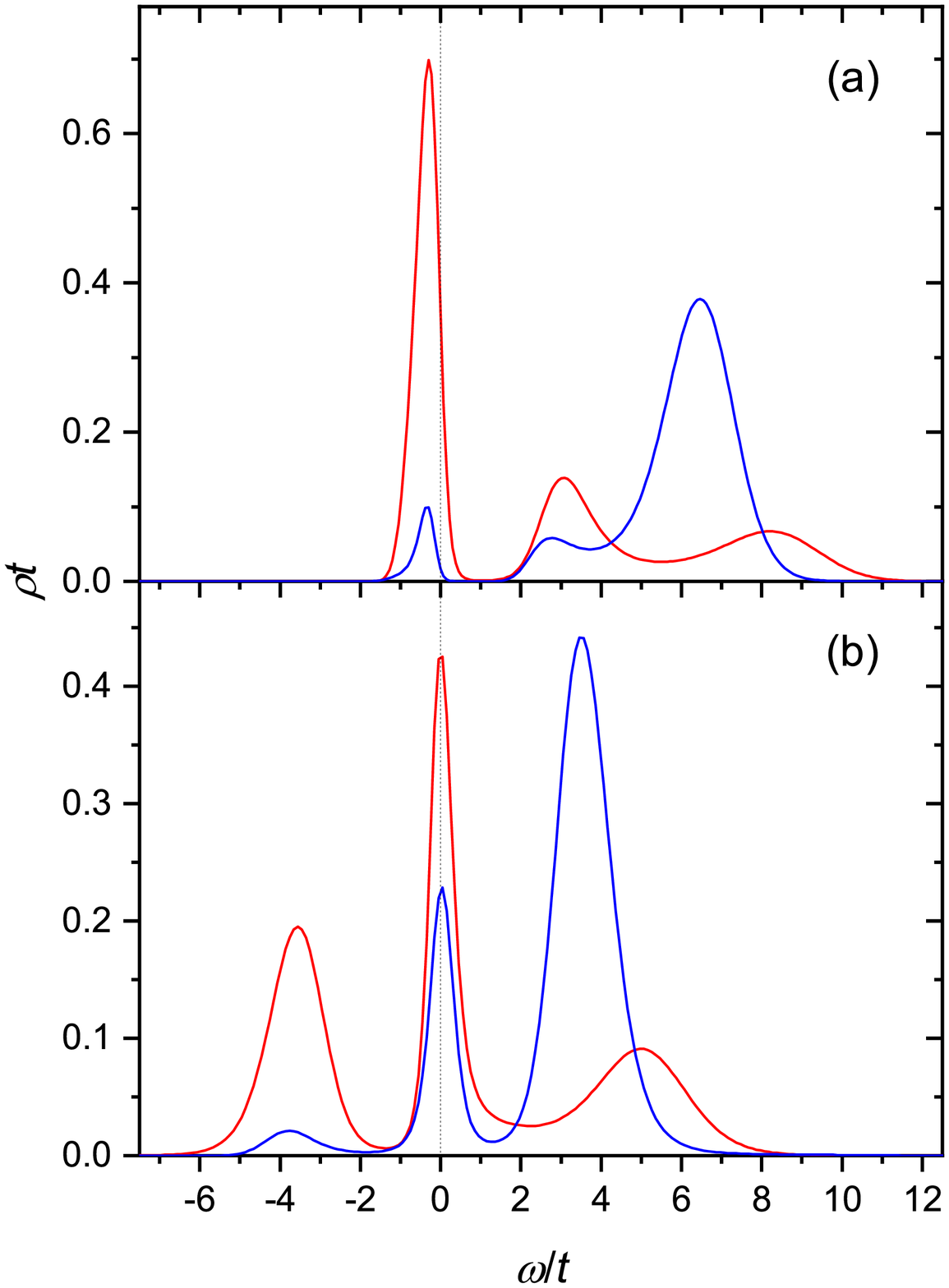}}}
\caption{The densities of states on copper (red lines) and oxygen (blue lines) sites, calculated for $U=5.7t$, $\Delta=5.4t$, $T=0.31t$, $\mu=-0.25t$ ($x=0.86$, a) and $\mu=2.7t$ ($x=1.26$, b).}\label{Fig6}
\end{figure}
The spectral functions, $A_i({\bf k},\omega)=-\pi^{-1}{\rm Im}\,G_{ii}({\bf k},\omega)$ are shown in Fig.~\ref{Fig5} for momenta along the symmetry lines of the Brillouin zone. Sets of parameters correspond to the electron doping (panel a), to the undoped case $x=1$ (panel b), and to the hole-doped situation (panel c). Again one can see some similarities between Fig.~\ref{Fig5}(c) and the spectral function calculated in Ref.~\cite{Weber08} for similar parameters. As follows from panels (a) and (c), states in the vicinity of the boundary of the magnetic Brillouin zone, the lines $(\pm\pi,0)-(0,\pm\pi)$, make the main contribution in the spin-polaron peak. This differs slightly from the result for the one-band Hubbard model with hopping terms to the second- and third-neighbor sites. In that model, states near the $\Gamma$ point contribute to the peak also \cite{Sherman19a}. In the case of electron doping in panel (a), the input of copper sites is dominant, while in the hole-doped case in panel (c) copper and oxygen sites contribute nearly equally to the peak. In the next section, we shall see that it is not the only difference between the hole and electron doping.

Figure~\ref{Fig6} demonstrates DOSs calculated for a somewhat higher temperature. As in the one-band model \cite{Sherman19a}, this small elevation of $T$ leads to the disappearance of the spin-polaron peak. The shape of the DOS and its behavior with doping are similar to those in the Hubbard-I approximation.

\section{Susceptibility, double occupancy, and squared site spin}
\begin{figure*}[t]
\centerline{\resizebox{1.7\columnwidth}{!}{\includegraphics{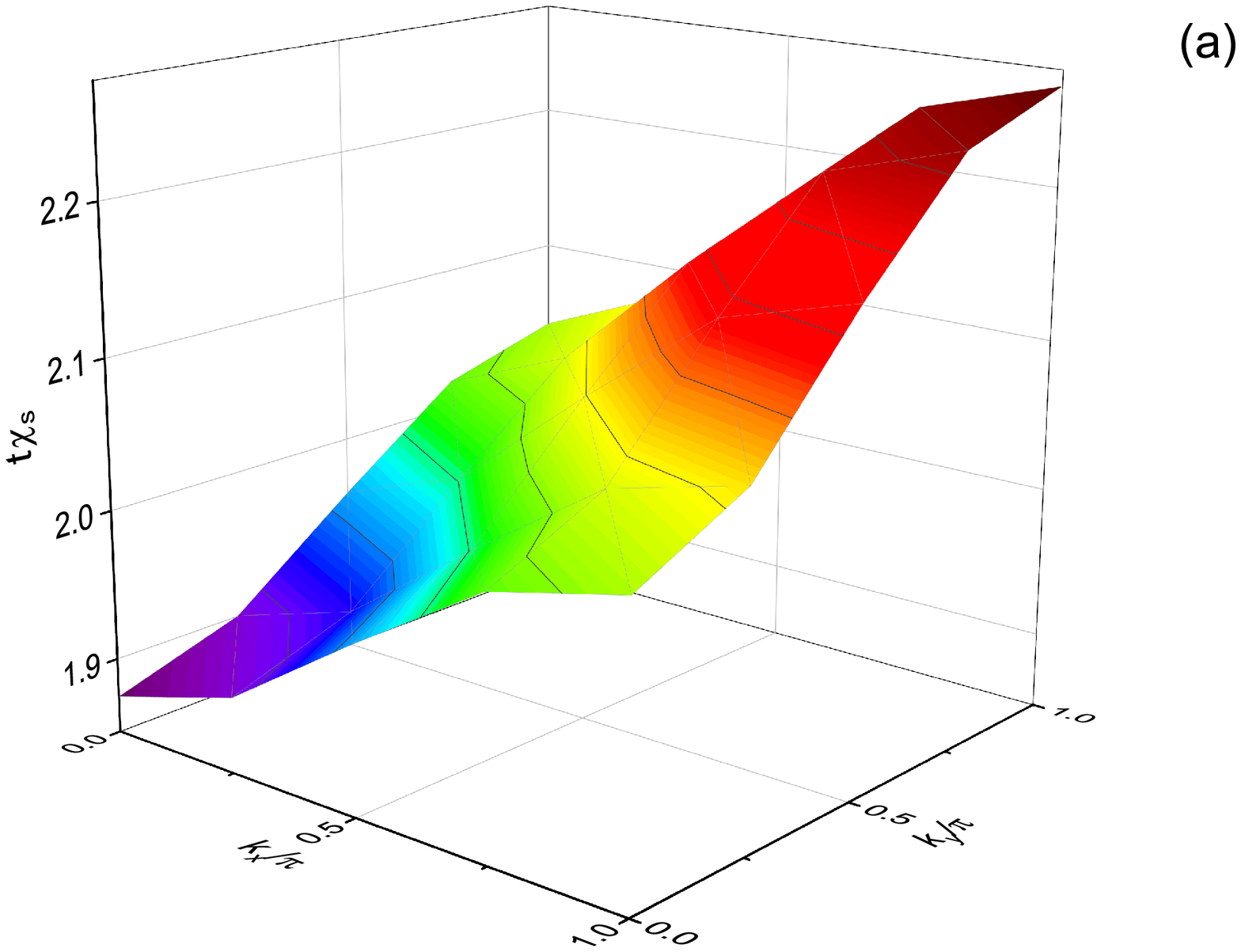}\hspace{5em} \includegraphics{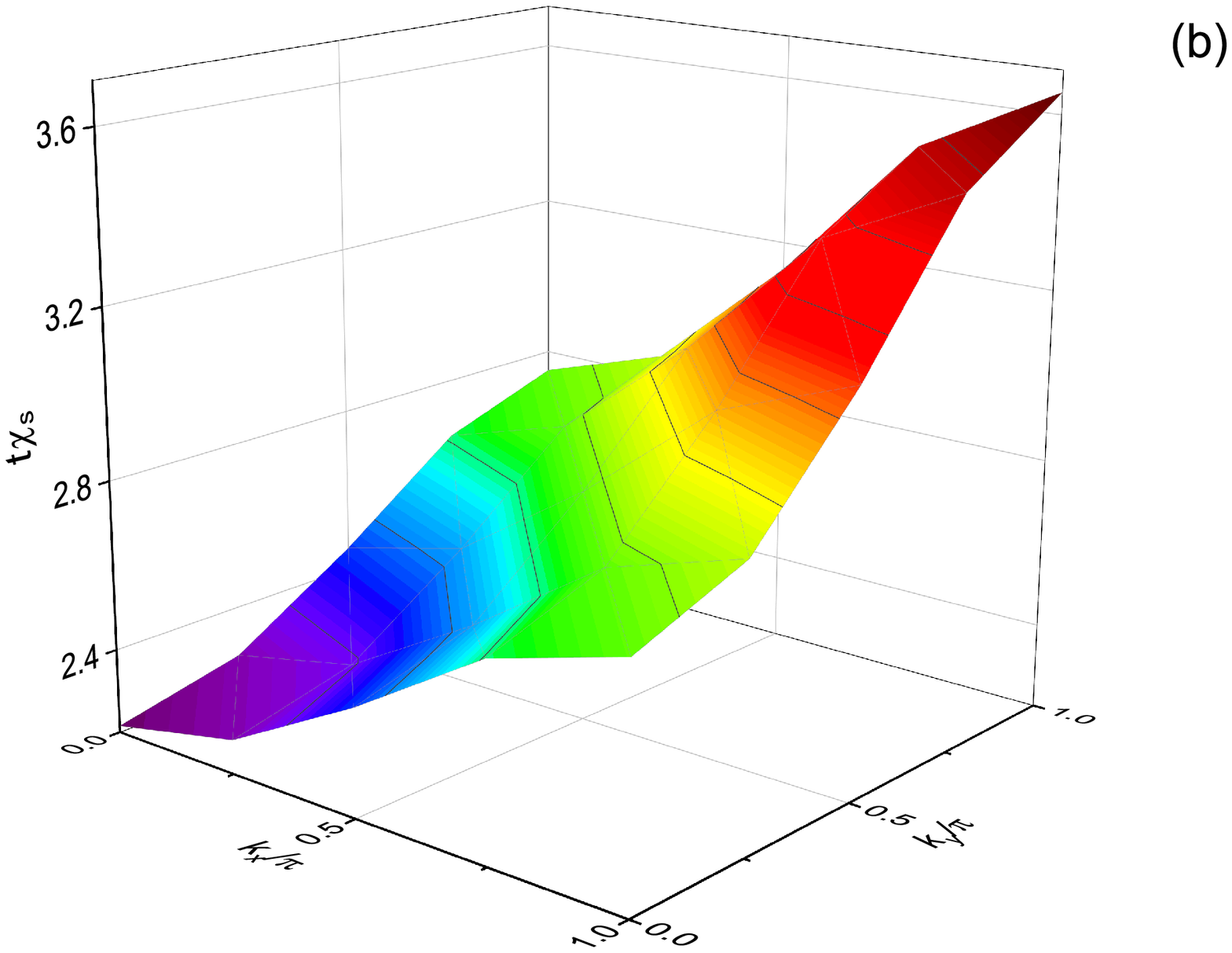}}}
\hspace{3ex}
\centerline{\resizebox{1.7\columnwidth}{!}{\includegraphics{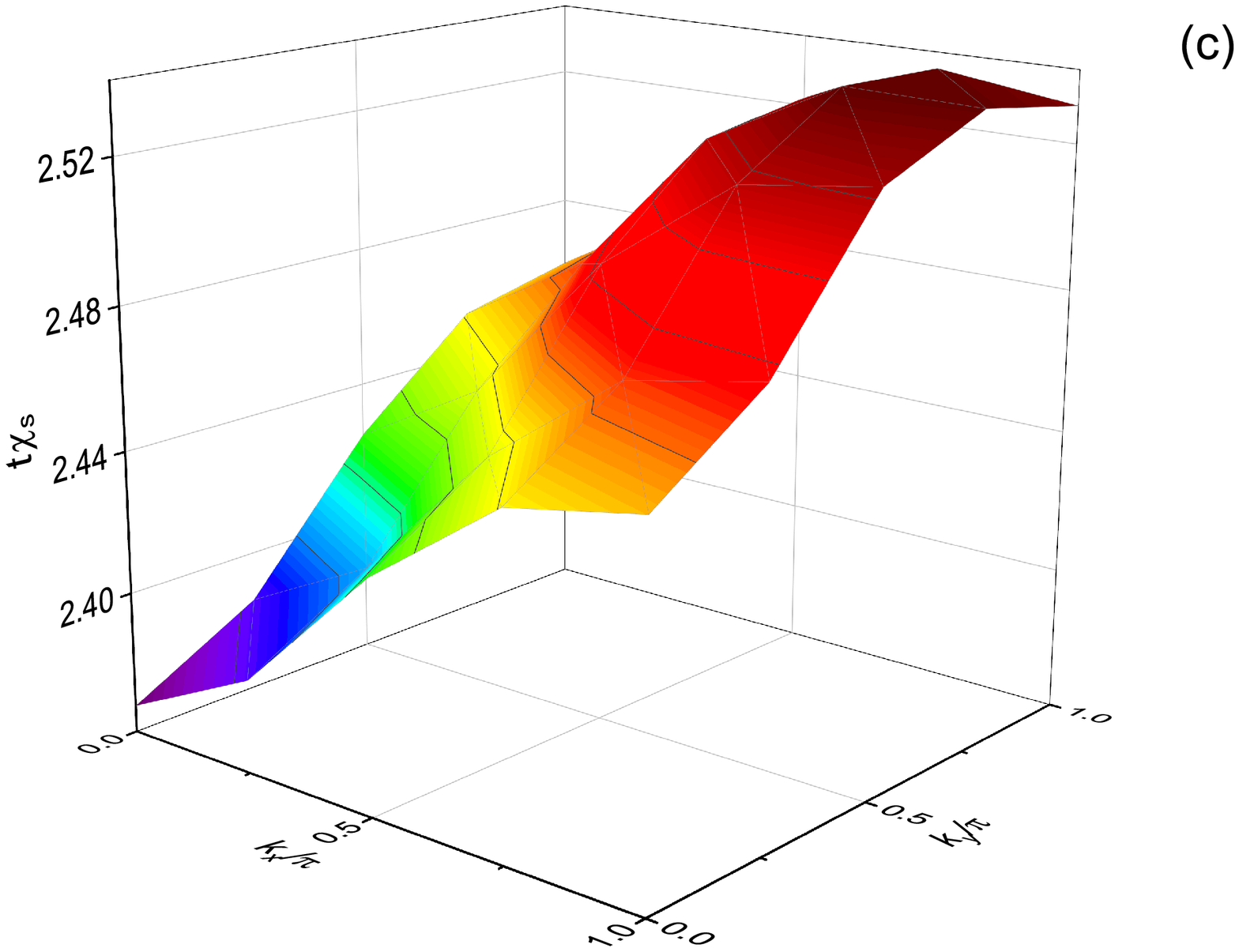}\hspace{5em} \includegraphics{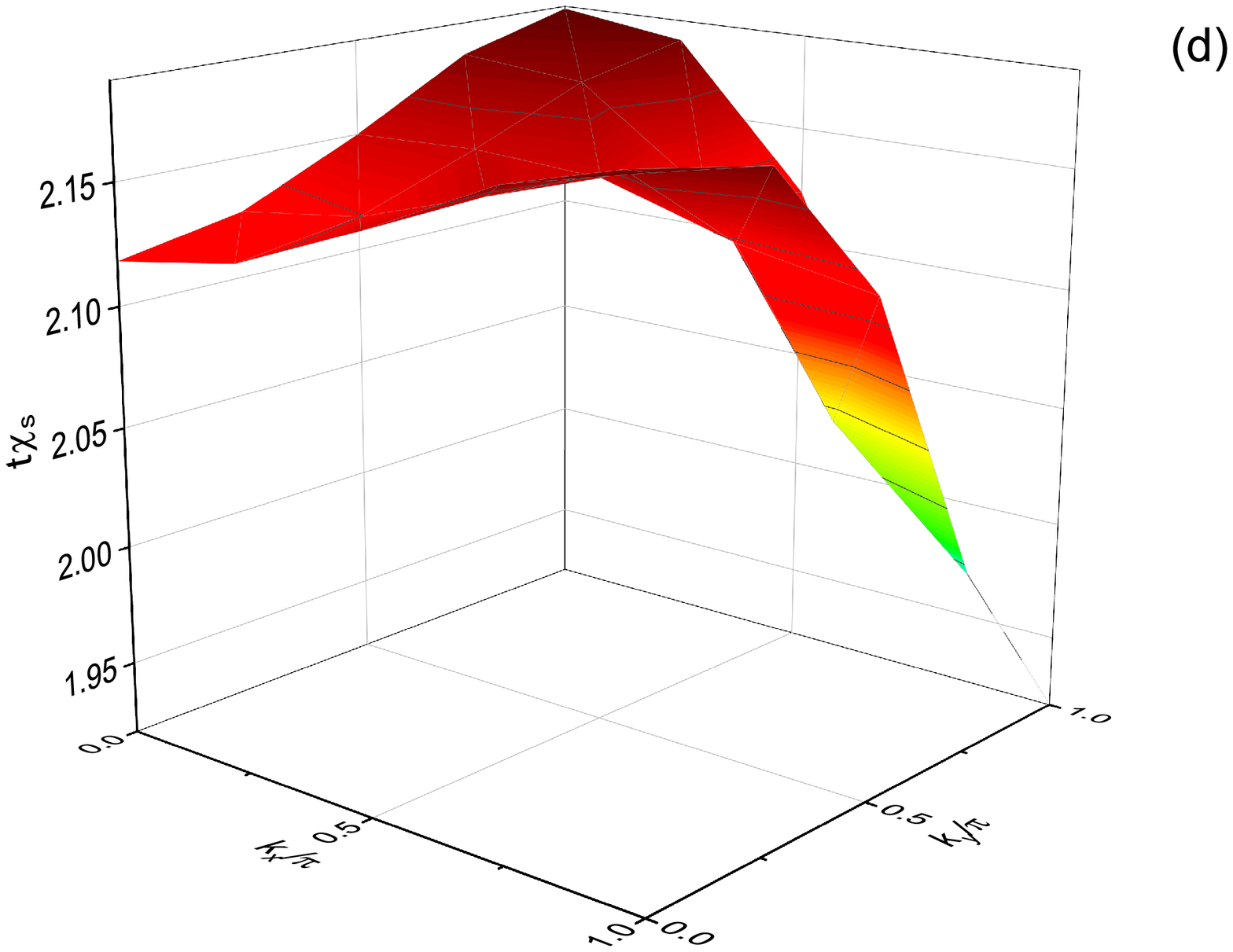}}}
\caption{The momentum dependence of the zero-frequency spin susceptibility in the first quadrant of the Brillouin zone. $U=5.7t$, $\Delta=5.4t$, $T=0.12t$, and $\mu=-0.5t$ ($x=0.77$, a), $t$ ($x=1$, b), $2.5t$ ($x=1.16$, c), $2.7t$ ($x=1.28$, d).} \label{Fig7}
\end{figure*}

The spin and charge susceptibilities,
\begin{eqnarray*}
\chi^{\rm sp}_{i'i}({\bf l'}\tau';{\bf l}\tau)&=&\langle{\cal T}\bar{a}_{{\bf l'}i'\sigma}(\tau') a_{{\bf l'}i',-\sigma}(\tau') \bar{a}_{{\bf l}i,-\sigma}(\tau) a_{{\bf l}i\sigma}(\tau)\rangle,\\
\chi^{\rm ch}_{i'i}({\bf l'}\tau';{\bf l}\tau)&=&\sum_\sigma\Big\langle{\cal T}\big[\bar{a}_{{\bf l'}i'\sigma'}(\tau') a_{{\bf l'}i'\sigma'}(\tau')-\langle n_{{\bf l'}i'\sigma'}\rangle\big]\\
&&\times\big[\bar{a}_{{\bf l}i\sigma}(\tau) a_{{\bf l}i\sigma}(\tau)-\langle n_{{\bf l}i\sigma}\rangle\big]\Big\rangle,
\end{eqnarray*}
can be calculated from the obtained Green's functions and vertices,
\begin{eqnarray}\label{suscept}
&&\chi^{\rm sp}_{i'i}({\bf k},\nu)=-\frac{T}{N}\sum_{{\bf k}j}G_{i'i}({\bf k+q},\nu+j) G_{ii'}({\bf q},j)\nonumber\\
&&\quad-T^2\sum_{jj'}\sum_{i_1\ldots i_4}F_{i'i_1i_4i'}({\bf k},\nu+j,j)\tilde{F}_{i_2iii_3}({\bf k},\nu+j',j')\nonumber\\
&&\quad\times V^a_{i_1i_2i_3i_4}({\bf k},\nu+j,\nu+j',j',j).
\end{eqnarray}
The expression for $\chi^{\rm ch}_{i'i}({\bf k},\nu)$ has the same form with $V^a$ substituted by $V^s$. In Eq.~(\ref{suscept}),
\begin{eqnarray*}
&&F_{ii_1i_2i}({\bf k},j',j)=\frac{1}{N}\sum_{\bf q}\Pi_{ii_1}({\bf k+q},j')\tilde{\Pi}_{i_2i}({\bf q},j'),\\
&&\tilde{F}_{i_1iii_2}({\bf k},j',j)=\frac{1}{N}\sum_{\bf q}\tilde{\Pi}_{i_1i}({\bf k+q},j')\Pi_{ii_2}({\bf q},j'),\\
&&\mbox{\boldmath $\Pi$}({\bf k},j)={\bf 1}+{\bf G}({\bf k},j){\bf t_k},\quad
\mbox{\boldmath $\tilde{\Pi}$}({\bf k},j)={\bf 1}+{\bf t_k}{\bf G}({\bf k},j).
\end{eqnarray*}

Calculations show that contributions of oxygen $\chi_{22}$ into zero-frequency susceptibilities are two orders of magnitude smaller than those of copper $\chi_{11}$. The exclusion is the case of heavy hole doping when $\chi^{\rm ch}_{22}$ becomes comparable to $\chi^{\rm ch}_{11}$. For the considered parameters, $\chi^{\rm ch}_{11}$ is an order of magnitude smaller than $\chi^{\rm sp}_{11}$. Therefore, the latter susceptibility will be considered below.

The momentum dependence of the susceptibility $\chi^{\rm sp}_{11}({\bf k},\nu=0)$ in the first quadrant of the Brillouin zone is shown in Fig.~\ref{Fig7}. For the undoped and electron-doped cases (panels b and a), the susceptibility is peaked at the wave vector $(\pi,\pi)$, which points to the short-range antiferromagnetic ordering. For hole doping, panels c and d indicate the appearance of the incommensurate magnetic ordering. The incommensurability parameter -- the distance between the momentum of the maximum of $\chi^{\rm sp}_{11}$ and $(\pi,\pi)$ -- grows with hole doping. Qualitatively similar behavior is observed in electron- and hole-doped cuprates \cite{Armitage,Fujita,Wilson,Damascelli}. Thus, the system described by the same Hamiltonian with the same parameters demonstrates the drastically different character of the magnetic response in the cases of the hole and electron doping.

When the zero-frequency spin susceptibility is peaked at $(\pi,\pi)$, its value at this momentum can serve as the measure of the proximity to the long-range antiferromagnetic ordering. Contrasting Fig.~\ref{Fig7}(b) with susceptibilities calculated in the one-band Hubbard model for half-filling and comparable parameters \cite{Sherman19,Sherman19a} one can see that $\chi^{\rm sp}({\bf k}=(\pi,\pi),\nu=0)$ is noticeably smaller in the present model. Thus, in the considered situation, in which at half-filling copper is spinful, and oxygen is spinless, the hybridization weakens spin correlations.

The asymmetry of the hole and electron doping is seen in the double occupancy and square of the site spin also. The double occupancy of the copper sites $D=\langle n_{{\bf l}1\sigma}n_{{\bf l}1,-\sigma} \rangle$ can be calculated from the equation
\begin{equation}\label{D}
D=\frac{T}{UN}\sum_{{\bf k}ji}{\rm e}^{{\rm i}\omega_j\eta}G_{1i}({\bf k},j)\Sigma_{i1}({\bf k},j),\quad\eta\rightarrow+0,
\end{equation}
where
\begin{equation*}
\mbox{\boldmath $\Sigma$}({\bf k},j)={\bf G}_0^{-1}({\bf k},j)-{\bf G}^{-1}({\bf k},j)
\end{equation*}
is the self-energy and
\begin{equation*}
{\bf G}_0({\bf k},j)=\left(\begin{array}{cc}
                            {\rm i}\omega_j+\mu & -2t\alpha_{\bf k} \\
                             -2t\alpha_{\bf k} & {\rm i}\omega_j-\Delta+\mu
                             \end{array}\right)^{-1}
\end{equation*}
is the unperturbed Green's function. Equation~(\ref{D}) is the generalization of the known expression for the one-band Hubbard model \cite{Vilk}. It is derived from the equation of motion for Green's function.

In the iteration procedure discussed in the previous section, we calculated Green's function for 60-80 Matsubara frequencies. To obtain $D$ from Eq.~(\ref{D}), we need ${\bf G}$ for much larger frequencies also. To calculate them, we used the asymptotics of Green's function,
\begin{eqnarray}\label{asympt}
{\bf G}({\bf k},j)&\stackrel{|j|\rightarrow\infty}{\longrightarrow}&\frac{1}{{\rm i}\omega_j}{\bf 1} -\frac{1}{\omega_j^2}\left(\begin{array}{cc}
                            Ux_1/2-\mu & 2t\alpha_{\bf k} \\
                             2t\alpha_{\bf k} & \Delta-\mu
                             \end{array}\right)\nonumber\\
&&+{\cal O}(\omega_j^{-3}).
\end{eqnarray}
This equation was derived by applying the Lehmann representation \cite{Abrikosov}. Green's function settles into this asymptote at frequencies smaller than the maximal frequencies calculated in the iteration.

\begin{figure}[t]
\centerline{\resizebox{0.9\columnwidth}{!}{\includegraphics{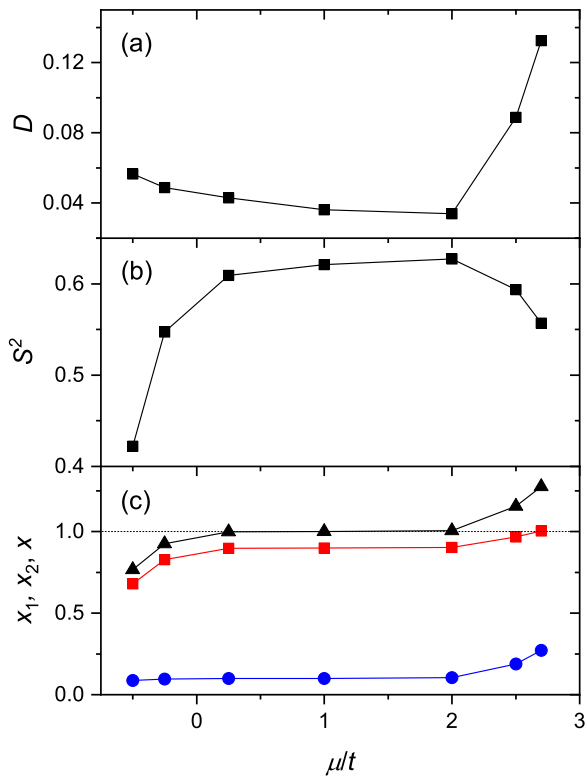}}}
\caption{(a) The double occupancy $D$, (b) squared spin on copper sites $S^2$, and (c) hole concentrations on oxygen (blue circles and line) and copper (red squares and line) sites as well as the total concentration (black triangles and line) as functions of the chemical potential. $U=5.7t$, $\Delta=5.4t$, and $T=0.12t$.}\label{Fig8}
\end{figure}
The square of the spin on copper sites is calculated from the equation
\begin{equation}\label{S2}
\left\langle{\bf S}^2_{{\bf l}1}\right\rangle=\frac{3}{4}x_1-\frac{3}{2}D.
\end{equation}
Results of calculations using Eqs.~(\ref{D})--(\ref{S2}) are shown in Fig.~\ref{Fig8}. The figure shows the dependencies of hole concentrations on $\mu$ also to indicate the boundaries of regions of the hole and electron doping, and to estimate the influence of the first term in Eq.~(\ref{S2}) on the behavior of $\left\langle{\bf S}^2_{{\bf l}1}\right\rangle$. As seen from the figure, the dependencies differ significantly in the hole- and electron-doped regions. The double occupancy remains as low as in the undoped case for large electron concentrations, while it grows rapidly with $\mu$ in the hole-doped case. The double occupancy can be considered as the measure of electron correlations. Based on this indicator, one can conclude that the system remains strongly correlated up to high levels of the electron doping, while correlations decay rapidly with the hole doping. Similar conclusions were also made from the analysis of experimental data \cite{Armitage}. On the contrary, $\left\langle{\bf S}^2_{{\bf l}1}\right\rangle$ decreases more rapidly in the electron-doped case. This behavior is connected with the fall of $x_1$ in the first term of Eq.~(\ref{S2}).

\section{Conclusion}
In this work, the two-band Hubbard model of cuprate CuO$_2$ planes, which contains $d_{x^2-y^2}$ copper and symmetric combinations of $p_\sigma$ oxygen orbitals, was investigated using the strong coupling diagram technique. In the Hubbard-I approximation, we considered the entire range of hole concentrations $0\leq x\leq 4$. It was shown that there exist five regions of the chemical potential, in which densities of states vary significantly, being related to different states. In a more elaborate approach, we take into account the interactions of holes with spin and charge fluctuations of all ranges. Derived equations for the hole Green's function were self-consistently solved for the set of parameters corresponding to cuprates. The main consequence of the interactions with spin and charge fluctuations is in the appearance of sharp maxima in the density of states and spectral functions at the Fermi level in permitted bands. Analogous maxima were observed in the one-band Hubbard model. They were related to the bound states of holes with spin excitations, which, by analogy with excitations in the $t$-$J$ model, were called spin polarons. They exist only at low temperatures. Polaron peaks are most intensive near the boundary of the magnetic Brillouin zone. In the case of electron doping, the peaks are mainly visible in spectral functions on copper sites, while for hole doping, their intensities are nearly equal in spectra on copper and oxygen sites.

Obtained Green's functions and vertices were used for calculating copper and oxygen contributions into the zero-frequency spin and charge susceptibilities. For the considered parameters, the copper contribution is much larger than that of oxygen sites. The exception is the case of a heavy hole doping when charge susceptibilities of both components are comparable. On copper sites, the spin susceptibility is much larger than the charge susceptibility. This spin susceptibility is peaked at the antiferromagnetic wave vector $(\pi,\pi)$ in undoped and electron-doped cases up to the concentration 0.23. Thus, in these conditions, the system demonstrates pronounced antiferromagnetic fluctuations. For the hole doping, the magnetic response becomes incommensurate, and the incommensurability parameter -- the distance between $(\pi,\pi)$ and the momentum of the susceptibility maximum -- grows with doping. This behavior of the model susceptibility coincides with peculiarities of the magnetic response in the hole- and electron-doped cuprates. The obtained results show a noticeable weakening of spin correlations in comparison with the one-band Hubbard model for comparable parameters.

The calculated double occupancy and square of the spin on a copper site demonstrate the sharp distinction
between the electron and hole doping also. Up to the electron concentration 0.23, the double occupancy remains as small as in the undoped case. This fact points to strong electron correlations, which retain in cuprates with a considerable excess of electrons. In contrast, the double occupancy grows rapidly with hole doping, indicating the decay of electron correlations. These results are in agreement with conclusions made from the analysis of experimental data in cuprates. The square of the site spin decreases more rapidly at the electron doping, which is related to the combined action of changes in the double occupancy and the hole concentration on copper sites.

The developed approach is applicable for any multi-band Hubbard model, allowing one to take into account the interactions of carriers with spin and charge fluctuations of all ranges. The derived expression for the second-order cumulant, Eq.~(\ref{C2}), can be used for this purpose. In particular, the approach can be applied to models describing iron-based high-temperature superconductors, for which there is firm evidence of strong correlation effects playing a key role in their unusual properties (see, e.g., \cite{Stadler}).

\end{document}